\documentclass[onecolumn]{aa}
\usepackage{graphicx}

\newcommand{\etal}{et\,al. }
\newcommand{\eg}{e.\,g. }

\begin{document}

\title{A difference boosting neural network for automated 
star-galaxy classification}
\titlerunning{Difference boosting neural network}
\authorrunning{N. S. Philip \etal}
\author{Ninan Sajeeth Philip\inst{1}
\and Yogesh Wadadekar\inst{2,3}
\and
 Ajit Kembhavi\inst{3}
\and
K. Babu Joseph\inst{1}}

\institute{Cochin University of Science and Technology, Kochi - 682
022, India \and Institut d'Astrophysique de Paris, 98bis Boulevard Arago, 75014 Paris, France \and Inter University Center for Astronomy and 
Astrophysics, Post Bag 4, Ganeshkhind, Pune - 411 007, India}

\offprints{Yogesh Wadadekar, \email{wadadeka@iap.fr}}


\abstract{
In this paper we describe the use of a new artificial neural network,
called the difference boosting neural network (DBNN), for automated
classification problems in astronomical data analysis. We illustrate
the capabilities of the network by applying it to star galaxy
classification using recently released, deep imaging data. We have
compared our results with classification made by the widely used
Source Extractor (SExtractor) package.  We show that while the
performance of the DBNN in star-galaxy classification is comparable to
that of SExtractor, it has the advantage of significantly higher speed
and flexibility during training as well as classification.
}

\maketitle

\section{Introduction}

Automated classification using neural networks has recently found
several applications in astronomy. These range from classification of
stellar (\eg Singh \etal 1998) and galactic (Folkes \etal 1996) spectra,
differentiating between stars and galaxies from imaging data
(\eg Odewahn \etal 1992), to detecting defects on CCDs (Rogers \&
Riess 1994).

Two powerful neural network paradigms (and their variants) have been
popular in the astronomical community for addressing classification
problems. These are, the Back-propagation algorithm (see McClelland \&
Rumelhart 1982) and the Self-Organizing Map (SOM; Kohonen 1982). The
basic difference between the two is that SOM can organize itself into
clusters unsupervised, meaning that no prior definition of the correct
output for an input vector is needed.  On the other hand,
back-propagation requires a carefully selected training set for
successful convergence. Both approaches have their own advantages and
disadvantages depending upon the application. When an input is
presented, SOM tries to associate it with a model vector that has
features that best match with those of the input vector. If no such
model vector is found, SOM assigns a new label to the object and
considers it as the model vector of a new class of objects. SOM is
therefore good in situations where the user does not know the possible
classes of objects {\it a priori}. However, SOM has the disadvantage
that it often classifies objects into different groups when in fact
they belong to a single group. In the astronomical context, this
often happens due to time-dependent variations in  features
associated with the data, \eg variations in seeing conditions.
Nevertheless, a number of variants of SOM have appeared with
different adaptation rules, distance measures and structures of the
map interconnections (Kohonen 1995). 

Back-propagation, on the other hand, is most suitable in situations
where one knows {\it a priori} the possible classes that an object can
belong to, and wants to classify objects with  a minimum error in the
classification. From a computational perspective, back-propagation is a
kind of function fitting process in which the parameters, generally
known as the connection weights, are modified so that the network is
able to map the system in terms of a nested sum of products of its
node functions, which  are generally sigmoid functions. The
back-propagation network requires a training set, a test set and often a
validation data set. In the training process the connection weights are
updated in proportion to the negative gradient of the error surface. The
network passes through the same set of data several times, lowering
the overall error in each pass. Since in many astronomical classification
problems the human observer is able to classify sources into their
possible classes directly, and only requires the machine to behave as a
faster classifier, back-propagation and its variants have been widely
used for classification of astronomical images and spectra. Another
advantage of the back-propagation network over SOM is that the user
has full control over the relevant parameters to be used for the
classification, and can therefore prevent the system from being 
misled by irrelevant
features in the training data. Raw data is preprocessed to extract the
significant parameters which then become the input for the neural
network.

Automated and accurate classification of objects into stars and
galaxies from optical (and near infrared) imaging data is an issue of
considerable interest. Artificial neural network based approaches to
the star galaxy classification problem include SOM (Miller \& Coe
1996), decision tree induction (Weir, Fayyad \& Djorgovski 1995) and
back propagation, which is the basis for SExtractor,  a widely used tool for
star-galaxy separation (Bertin \& Arnouts 1996).  One of the drawbacks of classification tools
such as SExtractor that employ back propagation is that it is
difficult to modify them for specific needs. With the advent of a
number of visible and near-infrared sky surveys, with varying
sensitivities and observing conditions, it is often desirable for the
user to be able to specify the optimum parameters for classification
after experimenting with several possibilities.  Trying new parameters
on a general purpose tool like SExtractor is computationally very
time consuming.  The objective of this paper is to present a tool that can
do the training at a faster rate, so that the network can be 
rapidly tested
with different sets of classification parameters to discover the
optimum parameter set.  The short training time makes the network 
very flexible and its potential astronomical applications are much 
wider than the star-galaxy classification problem discussed here.  

\section{The Difference Boosting Neural Network (DBNN)}

DBNN is based on a powerful and intuitive procedure originally
developed and used by mathematicians (Laplace 1812) for sensible
classification of objects. Named after its inventor
\cite{Bayes63}, Bayes' theorem, according to Laplace, is the
mathematical expression of common sense. Bayes' theorem computes the
conditional probability for the occurrence of an event, given that
another event which could lead to this event has occurred (for an
introduction to Bayes' theorem see Loredo 1990). In complex problems,
computation of the Bayesian probability becomes a laborious process.
However, a variant of the Bayesian classifier known as the
\textit{naive} Bayesian classifier (Elkan 1997) is able to compute the
Bayesian probability with reasonable accuracy under the assumption
that, given the class, the attribute vectors (arrays of 
parameter values) are independent.
However, in practice this independence assumption is frequently invalid and the
performance of the network degrades when there are correlated
attribute vectors.
 
The Difference Boosting Algorithm (DBNN) (Philip \& Joseph 2001) is
a computationally less intensive Bayesian classifier algorithm than
its peers, and is closely related to the naive Bayesian
classifier.  DBNN, however, does
not strictly follow the independence of attributes as a basic
criterion and allows some correlation between the attributes. It does
this by associating a threshold window with each of the attribute
values of the sample. The threshold window demands that all the
attribute values be in the range specified by the training set for
each class of the sample. When any of the attribute values is outside
the range specified by the threshold function, the confidence in the
classification is penalized by a certain factor.

\begin{table}
\caption{Output states of the XOR gate}\label{tab:Tab1}
\begin{tabular}{lllll}
\hline \hline
&Input 1 & Input 2 & Output \\ \hline
&0 & 0 & 0 \\  \hline
&0 & 1 & 1 \\  \hline
&1 & 0 & 1 \\  \hline
&1 & 1 & 0 \\  \hline
\end{tabular}
\end{table}

A popular example manifesting the correlation between input parameters
is the XOR gate. A typical XOR gate has two digital inputs and one
digital output. A digital state has only two possible values
represented as high or low state.  The XOR gate behaves such that its
output is a high only when the two inputs are dissimilar, as shown in
Table \ref{tab:Tab1}.  Although the conditions appear to be simple,
this is a case where the conditional independence is violated. For
example, from the knowledge of the value of only one of the inputs, it
is not possible to have any preferred knowledge about the class of the
object.  The actual class can be assigned only when both the inputs
are known together.  Since only the value of one of the inputs is used
by the naive Bayesian classifier at a time, it is not able to produce
a confidence level better than 50\% (both alternatives equally likely)
on such data. However, since DBNN takes into account the values of the
other parameters by the use of the window function, it is able to give
an accurate representation of the output states.  We then say that the
network has learned the XOR problem.

Some other advantages of the DBNN algorithm are its explicit
dependence on probability estimates, its ability to give an estimate
of the confidence value of a prediction and greater training speed. For the
particular application to star galaxy classification, the DBNN gives
good results with fewer input parameters than SExtractor.
Philip \& Joseph (2001), compared results from DBNN with the results
obtained by Schiffmann, Joost \& Werner (1994) on sixteen other
network models. While Schiffmann \etal (1994) report an average  
training time
of $\sim12$ hours on the dataset they used, DBNN on the same dataset
took only about 10 minutes for training. While their best result
from the eighteen models produced an accuracy of 98.48\% on
independent test data, DBNN gave an accuracy of 98.60\%.

One of the motivations for the DBNN classifier is that the human brain
looks for differences rather than details when it is faced with situations
that require distinction between almost identical objects. While the
standard Bayesian method is very elaborate and takes every possibility
into consideration, and the naive Bayesian ignores all possible
correlations between attribute values, DBNN is an attempt to have the
best of both worlds by highlighting only the differences.

\section{Working of the DBNN}

The working of the DBNN may be divided into three units. The first
unit computes Bayes' probability and the threshold function for each
of the training examples. The second unit consists of a gradient
descent boosting algorithm that enhances the differences in each of
the examples in an attempt to minimize the number of incorrectly
classified cases. At this stage, {\it boosting} is applied to the
connection weights for each of the probability components \( P(U_m \mid
C_k) \) of the attribute \(U_m\) belonging to an example from the
class \(C_k\). Initially all the connection weights are set to
unity. For the correctly classified object, the total probability
\(P(U \mid C_k)\), computed as the product of component probabilities
will be a maximum for \(C_k\), the class of the object given in the
training set. For the wrongly classified examples, for each of the
component probability values, the associated weights are incremented
by a factor \(\delta W_m\) which is proportional to the difference in
the total probability of membership of the example in the stated class
and that in the wrongly classified class. The exact value is computed
as $$
\delta W_m = \alpha \left( 1-\frac{P_k} {P^*_k} \right)
$$ {\noindent in a sequence of iterations through the training set.}
Here
\(P_k\) is the computed total probability for the actual class of the
object and \(P^*_k\) that of the wrongly classified class. The
parameter
\(\alpha\) is a learning constant functionally similar to the learning
constant in the back-propagation algorithm. It thus defines the rate
at which the algorithm updates its weight parameter. The third unit
computes the discriminant function (Bishop 1999) \( P(C_{k}\mid U) \)
as:

$$\label{eq3}
P(C_{k}\mid U)=\frac{\prod_{m}\hat{P}(U_{m}\cap C_{k}) 
W_{m}}{\sum_{j}\prod_{m}\hat{P}(U_{m}\cap C_{j}) W_{m}}.
$$
Here \(\hat{P}(U_m\cap C_k) \) stands for \(P(U_m\cap C_k)/P(C_k)\)
which from the axioms of set theory is equivalent to \(P(U_m \mid C_k)\).\\

In the implementation of the network, the actual classification is
done by selecting the class corresponding to a maximum value for the
discriminant function. Since this value is directly related to the
probability function, its value is also an estimate of the confidence
with which the network is able to do the classification. A low value
indicates that the classification is not reliable. Although a
network based on back-propagation also gives some probability estimates
on the confidence it has on a classification scheme, these are not
explicitly dependent on the probabilities of the distribution. Thus
while  such networks are vulnerable to divergent training vectors that
are invariably present in training samples, DBNN is able to assign low
probability estimates to such vectors. This is especially significant in
astronomical data analysis where one has to deal with variations in
the data due to atmospheric conditions and instrumental
limitations. Another significance of the approach is the simplicity in
the computation.  DBNN can be retrained with ease to adapt to the
variations in the observations enabling one to generate more accurate
catalogs.
 
In the following section, we describe the use of  the DBNN technique
to differentiate between stars and galaxies in broadband imaging
data. We chose to illustrate the capabilities of the DBNN by addressing
the star-galaxy classification problem for the following reasons:

\begin{enumerate}

\item A widely used benchmark implementation of the back propagation
algorithm is already available for tackling this problem in SExtractor.

\item High quality imaging data has recently become available from
ongoing optical surveys. The number of sources detected by such
surveys is large enough for us to construct moderately large training
and test sets from uniformly high quality data. The data we use here
is publicly accessible and our results can therefore be verified and
extended by other researchers.

\item Construction of a reasonably accurate training set is possible
from visual examination, given our experience with optical imaging data.

\end{enumerate}

\section{Separating stars and galaxies}

The star--galaxy classification problem addresses the task of labeling
objects in an image either as stars or as galaxies based on some
parameters extracted from them.  Classification of astronomical
objects at the limits of a survey is a rather difficult task, and
traditionally has been carried out by human experts with intuitive
skills and great experience. This approach is no longer feasible,
because of the staggering quantities of data being produced by large
surveys, and the need to bring objectivity into the classification, so
that results from different groups can be compared. It is thus
necessary to have machines that can perform the task with the
efficiency of a human expert (but at much greater speed) and with
robustness in the classification, over variations in observing
conditions.  

Processing the vast quantities of data produced by new and ongoing 
surveys and 
generating accurate catalogs of the objects detected in these
surveys is a formidable task, and reliable and fast classifiers are
much in demand.  Following the work by Odewahn \etal (1992), there has
been a growing interest in this area in the past decade.  SExtractor
(Bertin \& Arnouts 1996) is a popular, publicly available general
purpose tool for this application.  SExtractor accepts a FITS image of
a region of the sky as input and provides a catalog of the detected
objects as output. It has a built--in back propagation neural network
which was trained once for all by the authors of SExtractor 
using about $ 10^{6}$ simulated images of stars and
galaxies, generated under different conditions of pixel-scale, seeing
and detection limits. In
SExtractor an object is classified quantitatively by a {\it stellarity
index} ranging from zero to unity, with index zero representing a
galaxy and unity representing a star. The stellarity index is also a
crude measure of the confidence that SExtractor has in the classification. A
stellarity index of 0.0 or 1.0 indicates that SExtractor confidently
classifies these objects as galaxy and star respectively while a
stellarity index of 0.5 indicates that SExtractor is unable to classify the
object. The input to the neural network used by SExtractor
consists of nine parameters
for each object, extracted from the image after processing it through
a series of thresholding, deblending and photometric routines. Of the
nine input parameters, the first eight are isophotal areas and the
ninth one is the peak intensity for each object. In addition to these
nine parameters, a control parameter, the seeing full width at half
maximum (FWHM) of the image, is used to standardize the image
parameters against the  intrinsic fuzziness of the image due to the
seeing conditions. In practice, some fine
tuning of this control parameter is required for obtaining realistic
output from the network, due to the wide range of observing conditions
encountered in the data. A scheme for carrying out such tuning is
described in the SExtractor manual.

Among other packages proposed recently for star-galaxy classification
in wide field images is NExtractor (NExt) by Andreon \etal (2000).
NExt claims to be the first of its kind that uses a neural
network both for extracting the principal components in the feature
space, as well as for classification. The performance of the network
was evaluated over twenty five parameters that were expected to be
characteristic to the class label of the objects, and it was found that
six of these parameters, namely, the harmonic and Kron radius, two
gradients of the PSF, the second total moment and a ratio that
involves the measures of intensity and area of the observed object
were sufficient to produce optimum classification. A comparison of 
NExt performance with that of SExtractor by Andreon \etal (2000)
showed that NExt has a classification accuracy that is as good as or
better than SExtractor. The NExt code is not publicly available at the
present time (Andreon, personal communication) and a 
comparison with DBNN is not possible.

\subsection{Constructing the training set}

The first requirement for the construction of any good classifier is a
complete training set. Completeness here means that the training set
consists of examples with all possible variations in the target space
and that the feature vectors derived from them are distinct in the
feature space of their class labels. In the context of star--galaxy
classification, this means that the training set should contain
examples of the various morphologies and flux levels, of both stars
and galaxies, spanning the entire range of parameters of the objects
that are to be later classified.

We decided to construct our training set from an R band image from the
publicly available NOAO Deep Wide Field Survey (NDWFS). This survey
will eventually cover 18 square degrees of sky. The first data from
the survey obtained using the MOSAIC-I CCD camera on the KPNO 4m
Mayall telescope were released in January 2001. We chose to use data
from this survey because of its high dynamic range, large area coverage and
high sensitivity that allowed us to maintain uniformity between the
moderately large training set and numerous test sets. The training set
was carefully constructed from a subimage of the R band image NDWFSJ1426p3456 which has  the
best seeing conditions among the data currently released. Details of
the image are listed in Table \ref{tab:Tab2}.

\begin{table}
\caption{Summary of the NDWFS field used for constructing training and 
test sets}\label{tab:Tab2}
\begin{tabular}{ll}
\hline \hline
Field Name & NDWFSJ1426p3456\\
Filter & R\\
R.A. at field center (J2000)&14:26:01.41\\
Dec. at field center (J2000) &  +34:56:31.67\\
Field size & 36.960$\arcmin\times$38.367$\arcmin$\\ 
Total Exposure time (hours) & 1.2\\
Seeing FWHM (arcsec) & 1.16\\ 
\hline
\end{tabular}
\end{table}

We used SExtractor as a preprocessor for selection of objects for the
training set and for obtaining photometric parameters for
classification. The values of some critical configuration parameters
supplied to SExtractor for construction of the object catalog are
listed in Table \ref{tab:Tab3}. Saturated stars were excluded from the
training set by setting the SATUR\_LEVEL parameter. SEEING\_FWHM was
measured from the point spread function (PSF) of the brightest
unsaturated stars in the image. The DETECT\_MINAREA parameter was set
so that every selected object had a diameter of at least 1.8 times the
FWHM of the PSF. DETECT\_THRESH was set conservatively to 3 times the
standard deviation of the background which was estimated locally for
each source. ANALYSIS\_THRESH was set to a lower value to allow for
more reliable estimation of the classification parameters we used. No
cleaning or filtering of extracted sources was done. DEBLEND\_NTHRESH
and DEBLEND\_MINCONT were set by trial and error using the guidelines
in the SExtractor documentation. The following parameters were
obtained from descriptions of the NDWFS data products in the NOAO
archives - PIXEL\_SCALE, MAG\_ZEROPOINT and GAIN. SExtractor computes
several internal error flags for each object and reports these as the
catalog parameter FLAGS. Objects with a FLAGS parameter $\geq 4$ were
deleted from the training set. This ensured that saturated objects,
objects close to the image boundary, objects with incomplete aperture
or isophotal data and objects where a memory overflow occurred during
deblending or extraction were {\it not} used.

\begin{table}
\caption{Values of important SExtractor parameters used in 
construction of the
training and test sets}\label{tab:Tab3}
\begin{tabular}{ll} \hline\hline Parameter & Value \\
\hline 
DETECT\_MINAREA	&64\\
DETECT\_THRESH	& 3\\
ANALYSIS\_THRESH	&1.0\\
FILTER		& N\\	
DEBLEND\_NTHRESH	& 32\\		
DEBLEND\_MINCONT	& 0.01\\
CLEAN		&N\\
SATUR\_LEVEL	&49999.0\\
MAG\_ZEROPOINT & 30.698\\
GAIN	&46.2\\
PIXEL\_SCALE &0.258\\
SEEING\_FWHM &1.161\\
BACKPHOTO\_TYPE	&LOCAL\\		
THRESH\_TYPE&     RELATIVE\\
\hline
\end{tabular}

\end{table}

The training set was constructed from objects satisfying the above
criteria from a $2001\times2001$ pixel region of the image described
in Table \ref{tab:Tab2}. The image region we used was selected at
random. The objects were largely in the Kron-Cousins magnitude range
20--26. Objects brighter than this limit are mostly saturated stars
which were not used.  Each object in the training set was visually
classified as a star or galaxy by two of the authors working
separately, after examining the radial intensity profile, surface map
and intensity contours. Less than 2\% of the sources were
differently classified by the two authors. These discrepancies were
resolved by a combined examination by both authors. It was not
possible to visually classify 35 of the objects, and these were
deleted from the training set. All the deleted objects are fainter
than magnitude 26. Some details about the training set, named NDWF10,
 are given in Table \ref{tab:testset}. 
Visual classification of many of the brighter stars
was aided by the perceptibly non-circular PSF of the image. After
visual classification was complete, SExtractor classification for all
sources in the training set was obtained. 
An object-by-object comparision of the visual and
SExtractor classification showed that the latter was successful in
97.76\% of the cases in reproducing the results of the visual
classification (see Table \ref{tab:testset}).  
The number of stars in the training set is
considerably smaller than the number of galaxies because of the high
galactic latitude of the field and the faint magnitudes of objects in
the training set.  

\begin{table}
\caption{Details of the regions used in constructing the training
and test data sets.  Stars and galaxies from NDWF10 were used in
training the network, while the other two data sets were used to test 
the performance of the network after training. Each region was 
$2001\times 2001$ pixels in size.}\label{tab:testset}
\begin{tabular}{lllllllll}
\hline 
\hline
Data Label & R.A. (J2000) & Dec. (J2000) & Stars & Galaxies & Total \\
           &             &             &       &          &        \\
\hline     
NDWF10   & 14:26:28.76 & 34:59:19.94 & 83    & 319      & 402    \\
NDWF5     & 14:27:11.23 & 34:50:50.92 & 65    & 239      & 304   \\
NDWF14     & 14:26:28.18 & 35:07:55.69 & 89    & 319      & 408   \\
\hline
\end{tabular}

\end{table}

\subsection{Obtaining optimum parameters for classification}

Once a training set is available, the next task is to select the
parameters that the network will use for classification.  We tested
all available parameters extracted by SExtractor for their suitability
as classification parameters. We also derived some new parameters from
the basic parameters obtained from SExtractor. For the classification
we sought parameters which were (a)
{\it not} strongly dependent on the properties of the
instrument/telescope and on observing conditions; (b)  would not 
depend on photometric calibration of the
data, which is not always available; and (c) resulted in the 
clearest separation between stars and galaxies.  To meet the last
requirement, we plotted each parameter against the FWHM of the
intensity profile and identified the parameters which provided the
best separation. After extensive experimentation
with our training set data, we found that three parameters were most
suitable. These are: 

\begin{enumerate}

\item Elongation measure: This is the logarithm of the ratio of  second order
 moments along the major and minor axis of the lowest  isophote of the
object. For a star, the ratio should be near unity. For our training
set, this ratio is different from unity because of the slightly
elliptical PSF. 

\item The standardized FWHM measure: This is the logarithm of
the ratio of the FWHM of the object (obtained from a Gaussian fit to the
intensity profile) to the FWHM of the point spread function for the image.

\item The gradient parameter (slope): This is the logarithm of the ratio of the
central peak count to the FWHM of the object, normalized to the FWHM
of the point spread function for the image.

\end{enumerate}

\begin{figure*}
\centering
\includegraphics[width=17cm]{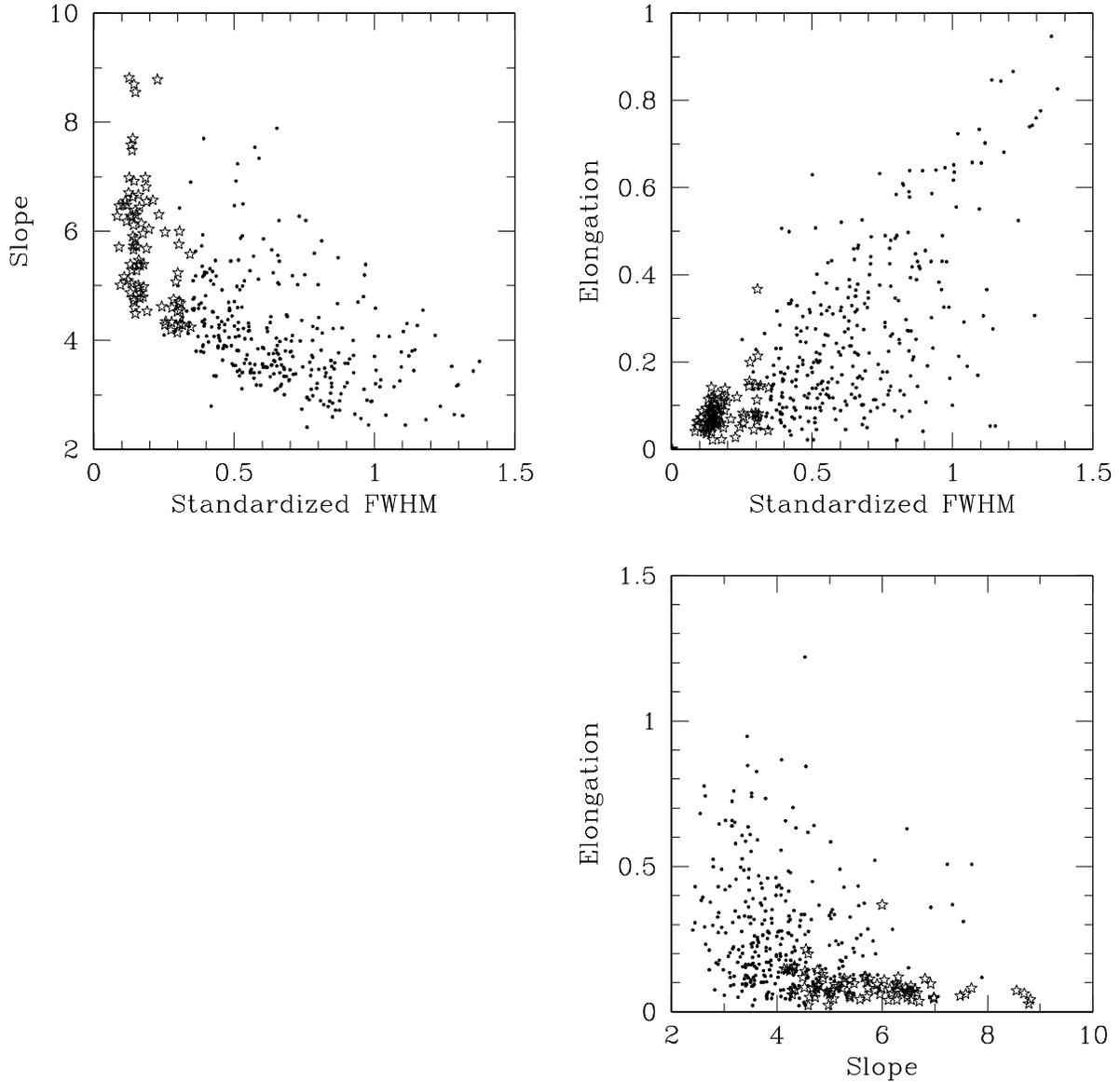} 
\caption{The figures show  clusters formed by stars and galaxies in the feature space. Galaxies are shown as dots and stars  as stars.}\label{fig:dbnn_fig}
\end{figure*}

We trained the  DBNN using the values
for these parameters for the visually classified set of stars and galaxies
as the training set. In Figure \ref{fig:dbnn_fig} we show plots of the three final 
DBNN parameters against each other, 
with stars and galaxies marked differently.  It is
clear that excellent separation between stars and galaxies is
obtained. 

\subsection{Testing the network performance \label{sec:res}}

We tested the network on 2 sub-regions (2001 $\times$ 2001 pixels each) of
the NDWFSJ1426p3456 field.  The central 
coordinates of the two test set images are
listed in Table \ref{tab:testset}. Using a different 
region of the same field for
testing ensures that erroneous classification due to variations in data
quality is not an issue. As in the case of the training set, these
sub-regions were also selected at random. The object catalogs for the
test sets were constructed using the same SExtractor configuration as
for the training set.  DBNN marked some objects as boundary examples,
meaning that their confidence level was not more than 10\% above the
plain guess estimate (50\%) regarding the class of the object.

In test set 1 (NDWF5), 32 out of 336 objects were deleted as they could
not be classified visually. Of the remaining 314 objects,
DBNN found 15 as marginal but
classified 10 of these correctly. Two objects were misclassified. In test
set 2 (NDWF14), 14 out of 422 objects were deleted for which visual
classification was not possible. Of the remaining,
DBNN marked 17 objects as marginal
but classified 12 of these correctly. One object was misclassified.  
The results for the two test sets are summarized in Table
\ref{tab:Tab5}. The
classification accuracy is marginally better than that of SExtractor.
The marginal superiority of DBNN, 
in the test set data, is not significant if some allowance is 
made for subjectivity in the construction of the test set. However, 
the fact that DBNN can obtain high classification accuracy with 
only 3 parameters as compared to 10 (9+ 1 control)  
parameters used by SExtractor is of some importance.

\begin{table}
\caption{Comparison of classification accuracy of the DBNN and
SExtractor on the NDWFS data.  There is no entry under DBNN for
NDWF10, since this data set was used in training the DBNN network.}
\label{tab:Tab5}
\begin{tabular}{lllll}
\hline \hline 
Data Label  & Classification Accuracy& Classification Accuracy\\
               & SExtractor &   DBNN \\
\hline     
NDWF10     &  97.76 \%  &      \\
NDWF5     &  96.05 \% & 97.70 \% \\
NDWF14      &  96.32 \%  & 98.52 \% \\
\hline
\end{tabular}

\end{table}

\subsection{Effects of image degradation}

An important consideration is to check the performance of DBNN (and
SExtractor) on low signal to noise images. In such images even visual
classification becomes difficult.  In order to examine the effects of
noise on the classification, we have therefore chosen to degrade the
training image NDWF10 by adding progressively higher levels of noise,
rather than use additional low S/N data. We have used the IRAF task
{\em mknoise} to increase the noise level of our training set. The
level of noise was controlled by using progressively higher values for
background counts. The original image has a background of 879
counts. Four additional images were created having a background count
of 20\%, 40\%, 60\% and 80\% of the original background. {\em mknoise}
was used to add Poisson noise to each of these 4 images, and they
represent progressively higher levels of background noise and lower
S/N ratio as compared to the original image. Note that the noise being
added by us here is in addition to the noise introduced during the
acquisition of the NDWFS data (which is already present in the original {\it undegraded} image). 
Sources are extracted from the degraded
images with the same SExtractor parameters used for the original
training set. The number of sources found in the degraded images are
listed in \ref{tab:degr_typ}.  As expected, the noisier the image, the
lower was the number of objects selected. The DBNN was {\em not}
retrained; sources in the degraded images were classified using the
DBNN trained with the original training set.

\begin{table}
\caption{Number of objects selected in the degraded images. The first line gives values for the undegraded data. The 4 degraded images are in decreasing order of S/N ratio. The criteria for object selection were the same as those for the undegraded training set image NDWF10}
\label{tab:degr_typ}
\begin{tabular}{llll}
\hline \hline
Image  & Background    &  Objects with &Number of objects\\
       &      &    $m_{R}> 25$ & Selected\\
\hline 
NDWF10 (undegraded image) &  879.0   &  313  & 402\\
NDWF104X5  & 175.8  & 313 &  402\\
NDWF103X5 & 351.6     & 2  & 49\\
NDWF102X5 & 527.4     & 1  &37\\
NDWF10X5 & 703.2     & 0  &30\\
\hline 
\end{tabular}
\end{table}

We have listed in Table \ref{tab:degrade} the performance of
SExtractor and DBNN on the degraded images. We find that DBNN
performance is slighter poorer than that of SExtractor on the fainter
sources.  This may be due to the fact that SExtractor uses the
magnitudes at 8 different isophotes as input parameters while DBNN
looks for gradients. For fainter objects, gradients are smaller,
making DBNN fail for a few faint objects.  A factor in
favour of DBNN is that it was trained with possibly contaminated
training data (due to limitations of the humans who constructed the
training and test sets) and can be retrained, while for SEx, the
training data was pristine (simulated) and frequent retraining is not
practical.

\begin{table}
\caption{Classification accuracy of SExtractor and DBNN as the NDWFS image
is gradually degraded. Objects that failed with a confidence level greater
than 60\% are marked as real failures. Number of objects with
$R$ magnitude greater than 25 in each set are shown in square
brackets}\label{tab:degrade}
\begin{tabular}{|l|l|l|l|l|l|l|l|l|}
\hline 
Image &\multicolumn {2}{c|} {Marginal Objects} & \multicolumn {2}{c|}{   Marginally Passed} & \multicolumn {2}{c|}  {Marginally Failed} & \multicolumn {2}{c|}{Real  failures} \\ \hline 

       &       DBNN &  SEx   &    DBNN &  SEx &  DBNN &  SEx &  DBNN &  SEx  \\
\hline 

NDWF10   &     31 & 34 & 21 [17]  &  31 [28]  &    10 [8] &    3 [3] &    0   &   6 [3]  \\
NDWF104X5 &    31 & 34 & 21 [17]  &   31 [28]  &    10 [8]  &   3 [3] &    0   &   6 [3] \\
NDWF103X5 &    4 & 3 &  3 [0]   &  3 [0]  &     1 [0]   &    0 &    0   &    1 [0] \\
NDWF102X5 &    5 & 2 &  2 [0]  &   2 [0]  &     3 [0]   &    0 &    0   &    1 [0]\\
NDWF10X5  &    1 & 1 &  0   &  1  &     1 [0]   &    0 &    1[0]   &    0 \\
\hline 
\end{tabular}
\end{table}

The second observation from the table is that, at brighter magnitudes,
DBNN produces more accurate classification on marginal objects
compared to SEx.  Also on objects that produce high confidence levels,
results from DBNN are marginally better than those of SEx. It is
important to keep in mind that the the confidence levels reported by a
neural network do not indicate the difficulty in visual classification by
humans. The confidence levels are parameter dependent and merely
quantify the appropriateness of a set of parameters. The actual
measures of efficiency of a classifier are (1) the total number of objects it
can classify with good confidence;  a good classifier should have
a minimum number of marginal objects at all magnitudes and (2) it should
produce minimum errors at high confidence levels.  The table shows
that DBNN does at least as well as SExtractor in overall efficiency of
classification on both these counts.

\section{Discussion and summary}

The results from the test data sets above indicate that the DBNN
classifier works as well as the widely used SExtractor software. All
instances of DBNN failure correspond to objects that do not have a
counterpart in the training set or objects that are difficult (but not
impossible) to classify visually. In both test sets, there is no
instance where an obvious misclassification has occurred.

As mentioned before, in addition to the consistency, the increase in
speed of the
training process is very significant here. Our training procedure for
DBNN on the 402 objects in the training dataset took 0.23 seconds on
an Intel Pentium III processor running at a clock speed of 700
MHz. Such short training times are invaluable when one has to
optimally deal with large datasets that are collected and processed
over a significantly wide span of time, demanding repeated retraining
of the classifier to account for variations in observing conditions
and use of newer and better parameters for classification. Data from
large surveys fall into this category.
Also, unlike BPNN, since DBNN is based on Bayesian
probability estimates, it is immune to diverse training vectors that
often appear in the training set due to noise in the observation. This
means that the performance of the network is likely to be
significantly better than the BPNN beyond the completeness limit.

In this paper we have illustrated the power of the technique by
applying it to the star galaxy classification problem. The technique
can easily be applied to all classification problems that currently
employ BPNN. For example, by using the large number of photometric and
spectroscopic parameters measured (for millions of objects) by surveys
such as the the Sloan Digital Sky Survey, it will be possible to apply
the DBNN technique to identify interesting samples for study in the
vast, largely unexplored parameter space. We are  in the process
of enhancing DBNN to solve problems that involve regression.

The source code, documentation and the training and test set 
images described in this paper may be downloaded from the URL:
http://www.iucaa.ernet.in/$\sim$nspp/dbnn.html

\begin{acknowledgements}

The authors would like to thank Ashish Mahabal for discussions and a
careful reading of the manuscript.  The first author would like to
express his sincere thanks to Inter University Center for Astronomy
and Astrophysics and the computer staff there for providing him all
the required facilities for the successful completion of this
project. We also thank the anonymous referee whose detailed
comments considerably improved this paper.

This work made use of images and data products provided by the NOAO Deep
Wide-Field Survey (Jannuzi and Dey 1999), which is supported by the National
Optical Astronomy Observatory (NOAO). NOAO is operated by AURA, Inc., under
a cooperative agreement with the National Science Foundation.

\end{acknowledgements}

{

}

\end{document}